\documentclass[prl,aps,onecolumn]{revtex4}
\usepackage{graphicx} 
\usepackage[usenames]{color}
\usepackage{amsmath,amssymb}
\usepackage{gensymb}
\usepackage{natbib}
\usepackage{xcolor}

\begin{document}

\title{Actuated rheology of magnetic micro-swimmers suspensions: emergence of motor and brake states}
\author{Benoit Vincenti$^{1}$, Carine Douarche$^{2}$, Eric Clement$^{1}$}
\affiliation{$^{1}$ Laboratoire de Physique et M\'ecanique des Milieux H\'et\'erog\`enes (PMMH), CNRS, ESPCI Paris, PSL Research University, Sorbonne Universit\'e, Universit\'e Paris Diderot, 10 rue Vauquelin, 75005 Paris, France\\
$^{2}$ Laboratoire de Physique des Solides, CNRS, Paris-Sud University, Paris-Saclay University, UMR 8502 – 91405 Orsay Cedex, France.}
\begin{abstract}
We study the effect of magnetic field on the rheology of magnetic micro-swimmers suspensions. We use a model of a dilute suspension under simple shear and subjected to a constant magnetic field. Particle shear stress is obtained for both pusher and puller types of micro-swimmers. In the limit of low shear rate, the rheology exhibits a constant shear stress, called \textit{actuated stress}, which only depends on the swimming activity of the particles. This stress is induced by the magnetic field and can be positive (\textit{brake state}) or negative (\textit{motor state}). In the limit of low magnetic fields, a scaling relation of the \textit{motor-brake} effect is derived as a function of the dimensionless parameters of the model. In this case, the shear stress is an affine function of the shear rate. The possibilities offered by such an active system to control the rheological response of a fluid are finally discussed.
\end{abstract}
\date{\today}
\maketitle

\section{I. INTRODUCTION}
Many micro-organisms are able to move autonomously in fluids at a very low Reynolds number \cite{Lauga2009} and recently, micron-size artificial particulate systems were designed to insure self-propulsion using either chemical \cite{Paxton2004, Maass2016}, magnetic excitations \cite{Dreyfus2005, Tierno2008}  or even the mixing of biological material with mechanical parts \cite{Martel2012, Williams2014}. The hydrodynamics of suspensions laden with such self-propelled objects is currently the focus of many fundamental studies \cite{Koch2011, Marchetti2013} and it has been found that original macroscopic constitutive properties can stem from the swimming activity of the suspended particles \cite{Hatwalne2004,Toner2005,Wu2000,Sokolov2009,Sokolov2010,DiLeonardo2010,Leptos2009,Rafai2010,Mino2011,Rusconi2014,Gachelin2014,Lopez2015,Jibuti2012}. According to the intrinsic nature of the propulsive mechanism, one can observe specific increases \cite{Rafai2010} (puller swimmers) or decreases (pusher swimmers) \cite{Sokolov2009,Gachelin2014,Lopez2015} of the viscosity with the swimmer concentration. Recent theory predicted an additional  "swimming pressure" contribution that will eventually contribute, at low shear rate, to lower the viscosity for both types of swimmers \cite{Brady2017}. Furthermore, it was found experimentally that in an intermediate range of concentrations, the macroscopic viscosity may even cancel in analogy with the superfluid transition \cite{Cates2008, Giomi2010,Lopez2015} of quantum liquids. These recent experimental results and the models proposed to account for them are reported in a review of fluid mechanics~\cite{Saintillan2018}.

In nature, some strains of bacteria are able to synthesize and assemble linear arrays of nano-magnets and have developed a biological sensitivity to the magnetic field direction \cite{Uebe2016,Reufer2014}. Such magnetotactic bacteria are able to move preferentially to one of the magnetic poles and are called accordingly north-seekers (NS) or south-seekers (SS). Recently, these suspensions were found to exhibit complex collective behaviors under flow and magnetic field~\cite{Waisbord2016}. 

In this article, we consider suspensions of motile elongated particles bearing an intrinsic magnetic moment along their swimming direction. In the simple magnetotactic model we present here, we do not consider any biological feedback on the swimming direction in response to the magnetic field. The magnetotactic sensitivity is only due to a passive alignment in the direction of the field. We are interested in understanding how the application of an external magnetic field can modify the macroscopic rheology of the suspension. The suspension is subjected to a simple shear and a constant magnetic field is applied at a given orientation with respect to the flow direction.  First, the swimming orientation distribution is computed in the framework of a Fokker-Planck equation that includes a stochastic disorientation process. Then, in this framework, we compute the particle-borne shear stress and establish, for any type of swimmer (pusher or puller), the emergence of new rheological states induced by the magnetic field and imputable only to the swimming activity of the particles. We finally discuss these results in the perspective of reproducing these specific states with magnetotactic bacteria or artificial micro-swimmers and using them to control the flow.\\

\section{II. MODEL OF MAGNETIC MICRO-SWIMMERS IN A SIMPLE SHEAR FLOW}

The active magnetic model we use consists of rod-shaped particles bearing a magnetic moment $\mathbf{m}=m\textbf{p}$ pointing in the swimming direction $\mathbf{p}$ (NS). The swimmer is described as an ellipsoidal slender rod of aspect ratio $r=L/a\gg 1$, where $L$ is its total length and $a$ its equatorial diameter. The swimming mechanism can either be of the \textit{pusher} or \textit{puller} type and its active hydrodynamic field is simplified as a force dipole of strength $\epsilon\sigma_0$ \cite{Lauga2009}, where $\sigma_{0}$ is a positive quantity and $\epsilon=1$ for \textit{pullers}, $-1$ for \textit{pushers}. We will only deal with dilute suspensions of number density $n$ such that $n a^{2}L \ll 1$. 

\begin{figure}
\centering
\includegraphics[scale=.380]{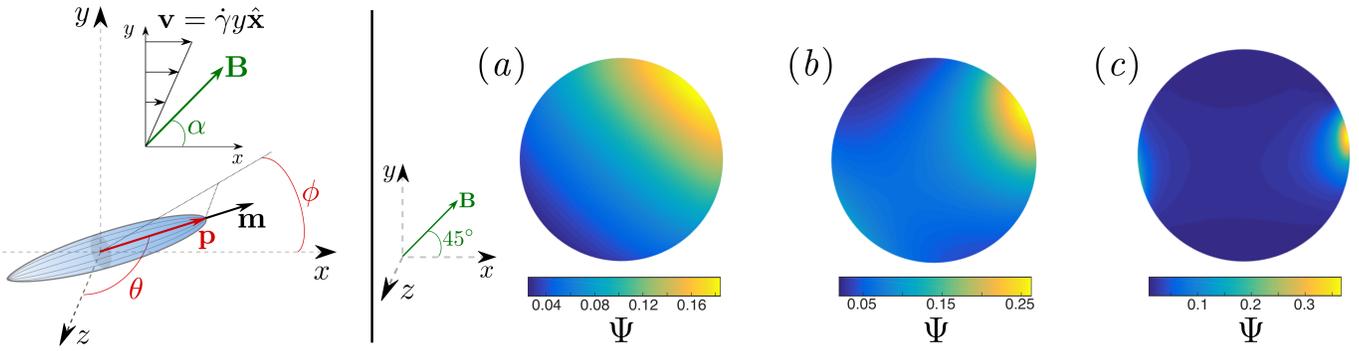}
\caption{\textit{Left} : 3D parameterization of a bacterium in spherical coordinates ($\theta$,$\phi$). The magnetic field $\mathbf{B}=B\mathbf{b}$ is contained in the ($x$,$y$) plane and its orientation in this plane is given by the angle $\alpha$. \textit{Right} : 3D representation of the orientation distribution function for $Pe_{m}=1$, $\alpha=45^{\circ}$ and $Pe_{H}=10^{-4}$ (\textit{a}), $Pe_{H}=4$ (\textit{b}) and $Pe_{H}=10$ (\textit{c}).}
\label{3D_para}
\end{figure}

A simple shear flow characterized by a velocity $\mathbf{v}=v_{x}\hat{\mathbf{x}}$ and a shear rate $\dot{\gamma}=\frac{\partial v_x}{\partial y}$ is applied to the suspension. The magnetic field $\mathbf{B}=B\mathbf{b}$ is oriented at an angle $\alpha$ from the flow direction $x$ (see figure~\ref{3D_para}). Importantly, because the shear and magnetic fields are spatially homogeneous, the orientation and spatial degrees of freedom are not coupled i.e. the orientation of each particle does not depend on space. In particular, only reorientation processes are important. These processes are described by a kinetic equation of the Jeffery-Bretherton type \cite{Jeffery1922,Bretherton1962}  that includes a magnetic  part due to the torque $\mathbf{m}\times \mathbf{B}$ \cite{Satoh2001}. Thus, the kinetic equation governing $\mathbf{p}$ is :
\begin{equation}
\mathbf{\dot{p}}=\dfrac{\mathrm{d}\mathbf{p}}{\mathrm{d}t}=(\overline{\overline{\mathbf{I}}}-\mathbf{p}\mathbf{p})(\beta \overline{\overline{\mathbf{E}}}+\overline{\overline{\mathbf{\Omega}}})\cdot\mathbf{p}+\mathbf{\Omega_{m}}\times\mathbf{p} \ ,
\label{kinematics}
\end{equation}
where the first term in the right-hand side of the equation stands for the flow contribution. $\overline{\overline{\mathbf{I}}}$ is the identity tensor, $\overline{\overline{\mathbf{E}}}=(1/2)\left(\nabla\mathbf{v}+(\nabla\mathbf{v})^{T}\right)$ is the strain-rate tensor and $\overline{\overline{\mathbf{\Omega}}}=(1/2)\left(\nabla\mathbf{v}-(\nabla\mathbf{v})^{T}\right)$ is the vorticity tensor. $\beta=(r^{2}-1)/(r^{2}+1)\simeq 1$ is the Bretherton parameter which will set to $\beta= 1$ from now on. $\mathbf{\Omega_{m}}=(mB/\xi_{r})\ \mathbf{p}\times\mathbf{b}$ is the rotation vector of the bacterium towards the magnetic field direction. The magnetic moment $\mathbf{m}$ of the particle relaxes towards the direction of the magnetic field with a characteristic time $\omega_{m}^{-1}=\xi_{r}/(mB)$, where $\xi_{r}$ is the rotational friction coefficient 
of the particle which can be computed in the framework of the slender body theory~: $\xi_{r}=\pi\eta_{0}L^{3}/\left(3\ln\left( 2L/a \right)\right)$.
 
Another source of swimming disorientation arises from a rotational diffusion term characterized by a coefficient $D_r$ which represents either a Brownian noise or the effect of a run and tumble process characterizing the bacterium motility. Under these hypothesis, the steady-state orientation distribution of the particle $\Psi(\theta,\phi)$ is solution of the following Fokker-Planck equation : 
\begin{equation}
\nabla_{s}\cdot (\mathbf{\dot{p}}\Psi)=D_{r}\nabla_{s}^{2}\Psi \ ,
\label{FP_equation_1}
\end{equation}
where $\nabla_{s}$ is the gradient operator on the unit sphere. The particle orientation is parameterized in spherical coordinates~: $\theta$ is the azimuthal angle while $\phi$ is the meridian angle (see figure~\ref{3D_para}). 

Equation~(\ref{FP_equation_1}) contains three non-dimensional parameters : the \textit{hydrodynamic rotational Peclet number} $Pe_{H}=\dot{\gamma}/D_{r}$~; the ratio of the rotational diffusion time to the magnetic relaxation time, $Pe_{m}=\omega_{m}/D_{r}$, which we call \textit{magnetic Peclet number} ; and $\alpha$, the magnetic field orientation. A detailed expansion of equation~(\ref{FP_equation_1}) in terms of these parameters reads~:
\begin{equation}
\nabla_{s}^{2}\Psi - Pe_{H}\Gamma_{\mathrm{shear}}(\Psi) + Pe_{m}\left[\cos(\alpha)\Gamma_{mx}(\Psi) +\sin(\alpha)\Gamma_{my}(\Psi) \right] = 0 
\label{developed_FP_eq}
\end{equation}
where $\Gamma_{\mathrm{shear}}$, $\Gamma_{mx}$ and $\Gamma_{my}$ are the following differential spherical operators :
\begin{equation}
\left\{
\begin{array}{l}
\Gamma_{\mathrm{shear}}(\Psi) = \dfrac{\sin(2\phi)}{2\sin\theta}\dfrac{\partial}{\partial\theta}\left( \sin^{2}\theta\cos\theta\Psi \right) - \dfrac{\partial}{\partial \phi}\left( \sin^{2}\phi \Psi\right)\\
\Gamma_{mx}(\Psi) = 2\sin\theta\cos\phi\Psi - \cos\theta\cos\phi\dfrac{\partial\Psi}{\partial\theta}+\dfrac{\sin\phi}{\sin\theta}\dfrac{\partial\Psi}{\partial\phi}\\
\Gamma_{my}(\Psi)=2\sin\theta\sin\phi\Psi-\cos\theta\sin\phi\dfrac{\partial\Psi}{\partial\theta}-\dfrac{\cos\phi}{\sin\theta}
\dfrac{\partial\Psi}{\partial\phi}
\end{array}
\right.
\end{equation}
Solving equation~(\ref{developed_FP_eq}) in 3D requires numerical tools. We used an expansion of $\Psi$ on a spherical harmonics basis (see Strand \textit{et al.} \cite{Kim1994} and Satoh \cite{Satoh2001} for technical details).  
The application of a magnetic field creates a preferential alignment of the particles in its direction in competition with both the alignment on the flow axis due to shear and the disorientation process due to the rotational diffusivity. For $Pe_{m}=1$, for which magnetic alignment is equivalent to diffusion disorientation, and orientation angle $\alpha=45^{\circ}$, we display on figure 1 the orientation distribution of the magnetic rod for $Pe_{H}= 10^{-4}$ (flow orientation negligible compared to magnetic orientation), $Pe_{H}= 4$ (equivalent contributions of flow and magnetic orientations), $Pe_{H}= 10$ (dominant flow orientation) to show the relative importance of the magnetic field compared to the flow orientation. While $Pe_{H}$ becomes important compared to $Pe_{m}$, the maximum of the orientation distribution becomes progressively aligned along the flow direction and the distribution becomes symmetric by rotation of $\pi$ around the $z$-axis, due to Jeffery orbits. \\
\indent Now we investigate the consequences of the swimming orientation distribution induced by the magnetic field on the mechanical response of the suspension. Following the approach of Saintillan \cite{Saintillan2010} (see also Haines et al. \cite{Haines2009}) who adapted the original method developed by Leal and Hinch~\cite{Leal&Hinch1976} for Brownian fibers, the total dimensional stress $\overline{\overline{\mathbf{\Sigma}}}$ can be expressed as a combination of both the fluid stress and the particle stress $\overline{\overline{\mathbf{\Sigma}}}_{p}$ :
\begin{equation}
\overline{\overline{\mathbf{\Sigma}}}=-P\overline{\overline{\mathbf{I}}}+2\eta_{s}\overline{\overline{\mathbf{E}}}+\overline{\overline{\mathbf{\Sigma}}}_{p} \ ,
\end{equation}
where $P$ is the fluid bulk pressure, $\eta_{s}$ is the suspending fluid dynamic viscosity. The particle stress contains four terms~\cite{Brenner1974, Jansons1983, Kim1994}:
\begin{equation}
\begin{array}{ll}
\overline{\overline{\mathbf{\Sigma}}}_{p}= & n\dfrac{\xi_{r}}{2}\left[<\mathbf{p}\mathbf{p}\mathbf{p}\mathbf{p}>-\dfrac{\overline{\overline{\mathbf{I}}}}{3}<\mathbf{p}\mathbf{p}>\right]:\overline{\overline{\mathbf{E}}}\\
& \\
& + 3nD_{r}\xi_{r}\left[ <\mathbf{p}\mathbf{p}> -\dfrac{\overline{\overline{\mathbf{I}}}}{3}\right]+\epsilon n\sigma_{0}\left[ <\mathbf{p}\mathbf{p}> -\dfrac{\overline{\overline{\mathbf{I}}}}{3}\right]\\
& \\
& - nmB<\mathbf{b_{\perp}p}>
\end{array}
\label{stress}
\end{equation}
where $\mathbf{b_{\perp}}=(\overline{\overline{\mathbf{I}}}-\mathbf{p}\mathbf{p})\cdot\mathbf{b}$ is the normalized projection of the magnetic field onto the plane perpendicular to the rod.

We will focus our investigation on $\overline{\overline{\mathbf{\Sigma}}}_{p}$. First, let us consider a dimensionless version of it : 
\begin{equation}
\widetilde{\overline{\overline{\mathbf{\Sigma}}}}_{p} =\overline{\overline{\mathbf{\Sigma}}}_{p}/ (n \sigma_{0}) 
\end{equation}
The energy density $n \sigma_{0}$ represents the maximal work per unit volume stemming from the swimming activity which is characterized microscopically by a time scale $t_{H}=\xi_{r}/\sigma_{0}$ needed for the swimmer to move the fluid over its own size. This time scale is used to define an \textit{activity number} : $\mathcal{A}=1/(D_{r}t_{H})$. The higher $\mathcal{A}$ the more directionally persistent is the bacterial swimming.
The ($x$,$y$) component of the dimensionless particle stress contains four terms and reads (see \cite{Brenner1974, Jansons1983, Kim1994} for details of the calculation) :
\begin{equation}
\begin{array}{ll}
\left(\widetilde{\overline{\overline{\mathbf{\Sigma}}}}_{p}\right)_{xy} = & \underbrace{\dfrac{1}{2}<p_{x}^{2}p_{y}^{2}>\dfrac{Pe_{H}}{\mathcal{A}}}_{\widetilde{\Sigma}_{\mathrm{drag}}}+\underbrace{\dfrac{3}{\mathcal{A}}<p_{x}p_{y}>}_{\widetilde{\Sigma}_{\mathrm{diff}}}\\
+\underbrace{\epsilon<p_{x}p_{y}>}_{\widetilde{\Sigma}_{\mathrm{act}}} & + \underbrace{\dfrac{Pe_{m}}{\mathcal{A}}<\left[p_{x}p_{y}b_{y}-b_{x}\left(1-p_{x}^2\right)\right]p_{y}>}_{\widetilde{\Sigma}_{\mathrm{mag}}}
\end{array}
\label{nondim_particle_stress}
\end{equation}
where $\widetilde{\Sigma}_{\mathrm{drag}}$ and $\widetilde{\Sigma}_{\mathrm{diff}}$ are passive contributions and account for drag on the surface of the particle from shear flow and diffusive process respectively; $\widetilde{\Sigma}_{\mathrm{act}}$ is related to the swimming activity of the particle; $\widetilde{\Sigma}_{\mathrm{mag}}$ represents the stress due to the perturbation of the flow by the magnetic field driven rotation of the particle. The brackets $<,>$ correspond to an angular average weighted by $\Psi$. Note that expression~(\ref{nondim_particle_stress}) is only valid for $\beta=1$. See section IV discussing the case of $\beta=0$.\\

\section{III. THE \textit{MOTOR-BRAKE} EFFECT}

$\left(\widetilde{\overline{\overline{\mathbf{\Sigma}}}}_{p}\right)_{xy}$ will be denoted $\widetilde{\Sigma}_{p}$. We restrict the investigations to Peclet numbers such that swimming remains the dominant contribution in the rheological response. Accordingly, in the following, we decided to fix the activity number at a value $\mathcal{A}=10$. 
\begin{figure}[t]
\centering
\includegraphics[scale=.395]{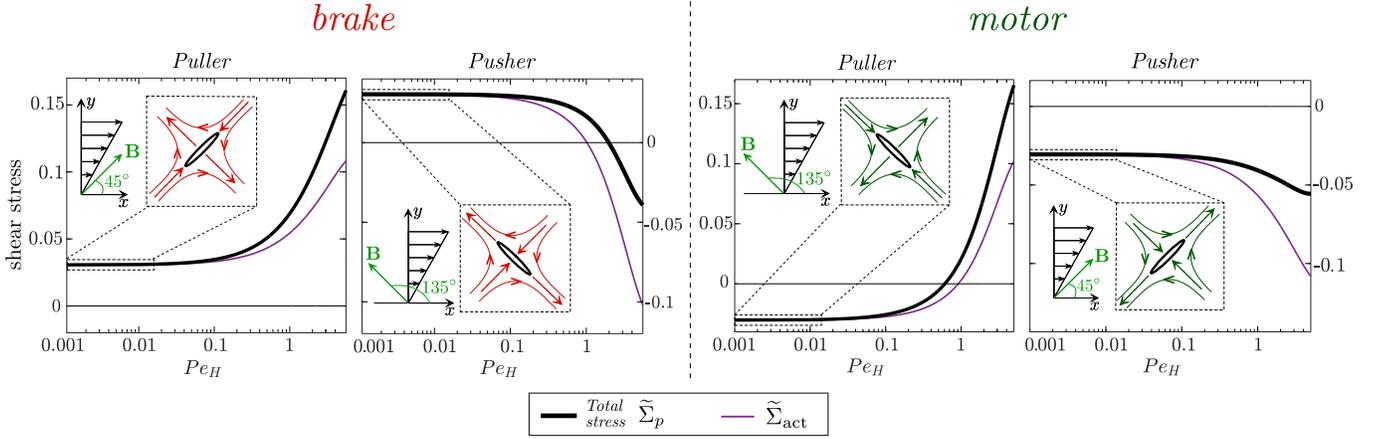}
\caption{Rescaled particle shear stress $\widetilde{\Sigma}_{p}$, derived numerically from equation~(\ref{nondim_particle_stress}), as a function of the rotational Peclet number $Pe_{H}$ ($\alpha=45^{\circ}$, $135^{\circ}$ ; $Pe_{m}=1$ ; $\mathcal{A}=10$ ; $\epsilon=1$ (\textit{puller}) and $-1$ (\textit{pusher})). The active contribution $\widetilde{\Sigma}_{\mathrm{act}}$ is color-labelled and referred to equation~(\ref{nondim_particle_stress}). When $Pe_{H}\ll 1$, $\widetilde{\Sigma}_{p}$ is equal to the active stress and tends to a constant value, the \textit{actuated stress}. Depending on the sign of the actuated stress, the suspension can be turned to \textit{brake} and \textit{motor} states. For each of the four graphs, the \textit{brake} and \textit{motor} states are interpreted by a sketch in which we show the orientation of the elongated particles relatively to the flow direction. The red (\textit{brake}) and green (\textit{motor}) lines and arrows correspond to the stream lines created by the hydrodynamic dipole of each particle (represented by a black ellipsoid).}
\label{break_motor}
\end{figure}
\begin{figure}[t]
\centering
\includegraphics[scale=.50]{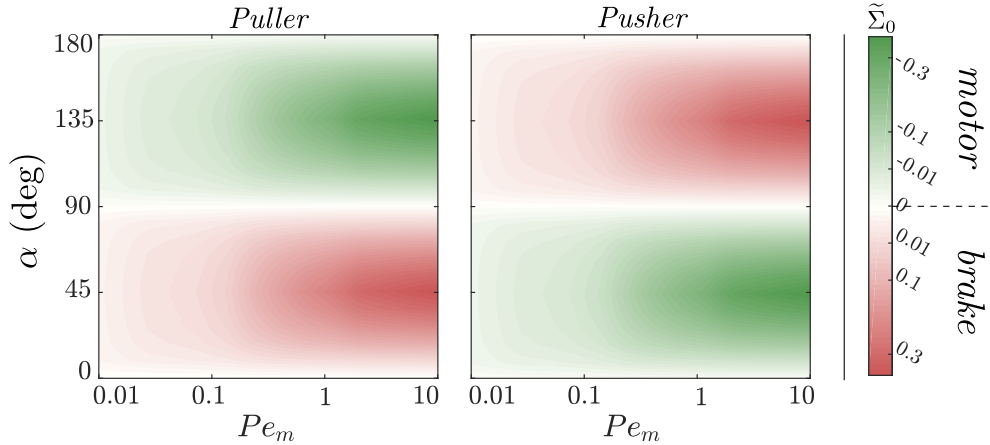}
\caption{Phase diagram in the ($Pe_{m}$, $\alpha$) space of the \textit{motor-brake} effect for \textit{puller} and \textit{pusher} swimmers. Positive values and negative values of the \textit{actuated stress} $\widetilde{\Sigma}_{0}$ are respectively red and green-labelled and correspond to \textit{brake} and \textit{motor} states. The relative magnitude of the actuated stress is indicated by a logarithmic color gradient.}
\label{phase_diagrams}
\end{figure}

\begin{figure}[h!]
\centering
\includegraphics[scale=.5]{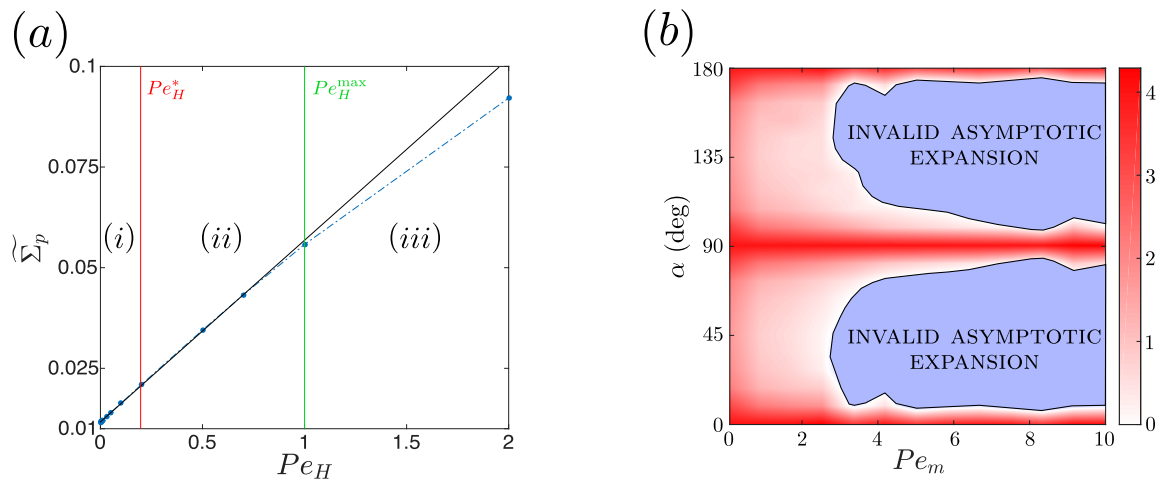}
\caption{(a). Example of relation $\widetilde{\Sigma}_{p} (Pe_{H})$ obtained from a full numerical solution of the problem for $\mathcal{A}=10$, $\alpha=45^{\circ}$, $Pe_{m}=0.6$ (symbols). The solid line is a linear fit taking into account the data up to the limit $Pe_{H}^{\mathrm{max}}$ such that the goodness of the fit $R^{2}$ remains larger that $0.999$. The slope of the curve yields $Pe_{H}^{*}$ and the intercept  $\widetilde{\Sigma}_{0}$. In \textit{regime (i)},  the \textit{actuated stress} dominates. In \textit{regime (ii)}, the stress is essentially linear in shear rate, the \textit{actuated stress} is negligible. In \textit{regime (iii)}, the affine expansion is no longer valid because other stress terms contribute to the shear stress. When the calculated $Pe_{H}^{*} > Pe_{H}^{\mathrm{max}}$, then we conclude that the regime $\widetilde{\Sigma}_{p}=\widetilde{\Sigma}_{0}+\eta_{p}Pe_{H}$ is never reached : the stress is constant at low $Pe_{H}$, up to a transition to a regime not anymore dominated by the active and drag stresses.\\
(b). Validity of the affine expansion of the stress~(\ref{asymptotic_particle_stress}) in the parameter space ($Pe_{m}$, $\alpha$) for $\mathcal{A}=10$. The color code corresponds to the value of $\Delta=\log_{10}(Pe_{H}^{\mathrm{max}})-\log_{10}(Pe_{H}^{*})$, which is the range of validity, in decades of $Pe_{H}$, for which equation~(\ref{asymptotic_particle_stress}) holds. When parameters are chosen in the regions in blue, equation~(\ref{asymptotic_particle_stress}) is no longer valid and other terms have to be taken into account in the expansion of the shear stress.}
\label{illustration_fit}
\label{diagramme_validite_Bingham}
\end{figure}

\begin{figure}[t]
\centering
\includegraphics[scale=.60]{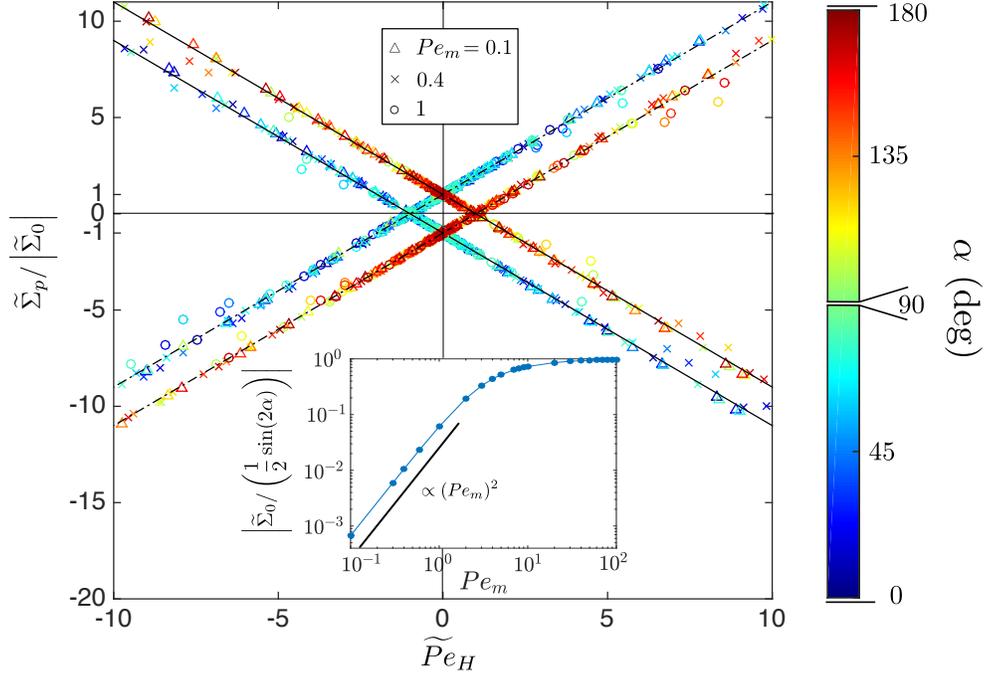}
\caption{Rescaled particle shear stress $\widetilde{\Sigma}_{p}/\lvert\widetilde{\Sigma}_{0}\rvert$ as a function of the rescaled Peclet number $\widetilde{Pe}_{H}$ for different values of $\alpha$, for $Pe_{m}=0.1$, $0.4$ and $1$ and for both \textit{pusher} and \textit{puller} swimmers. The $\alpha$ values are indicated by the color code. Values $\alpha=0^{\circ},90^{\circ},180^{\circ}$ are excluded because they correspond to the situation $\widetilde{\Sigma}_{0}=0$, for which the master equation~(\ref{equation_rescaling}) is not defined. The numerical solution displays, for a wide range of $Pe_{H}$ and $Pe_{m}$, a collapse onto expression (8), represented in black solid (pushers) and doted (pullers) lines. Negative values of shear rate correspond to a gradient of flow velocity in the -$y$ direction, and vorticity in the $z$ direction. By reversing the shear flow, the \textit{actuated stress} keeps its sign and the \textit{motor-brake} effect is reversed (from motor to brake state and \textit{vice-versa}). \textit{Inset~:} Rescaled actuated stress $\left\lvert\widetilde{\Sigma}_{0}/\left(\frac{1}{2}\sin(2\alpha)\right)\right\rvert$ as a function of the magnetic Peclet number $Pe_{m}$. It exhibits a scaling in $Pe_{m}^{2}$ for $Pe_{m}\ll 1$ as predicted analytically. For $Pe_{m}\gg 1$, it saturates.}
\label{collapse_plot}
\end{figure}

On figure~\ref{break_motor}, we display the behavior of $\widetilde{\Sigma}_{p}$ with respect to $Pe_{H}$ for different orientations of $\mathbf{B}$. An important feature of the rheological response is that for $Pe_{H} \ll 1$ and $Pe_{H}\ll Pe_{m}$ (for any $Pe_{m}$), the diffusive $\widetilde{\Sigma}_{\mathrm{diff}}$ and magnetic $\widetilde{\Sigma}_{\textrm{mag}}$ stresses do compensate each other. Then, the particle stress $\widetilde{\Sigma}_{p}$ is mainly a combination of the active $\widetilde{\Sigma}_{\mathrm{act}}$ and the drag $\widetilde{\Sigma}_{\mathrm{drag}}$ stresses. While the drag stress vanishes for $Pe_{H}\rightarrow 0$, the particle stress $\widetilde{\Sigma}_{p}$, which is completely determined by the active stress contribution, tends linearly to a non-zero constant (see figure~\ref{break_motor} for \textit{puller} and \textit{pusher} swimmers). This constant stress will be called the \textit{actuated stress} and denoted $\widetilde{\Sigma}_{0}$. It is created by the swimming activity and induced by the magnetic field. Indeed, at these low $Pe_{H}$, the magnetic particle is essentially oriented in the direction of the magnetic field and both \textit{pusher} and \textit{puller} swimmers can increase the shearing of the fluid (\textit{motor state}) or decrease it (\textit{brake state}). This is illustrated on figure 2 by the drawing of the swimmer orientations and a sketch of the corresponding flow lines. \\
\indent The intensity of the \textit{motor-brake} effect relies on the value of the \textit{actuated stress} $\widetilde{\Sigma}_{0}$ which itself depends on the magnitude and orientation of the magnetic field. On figure~\ref{phase_diagrams}, we plot two phase-diagrams in the ($Pe_{m}$, $\alpha$) space for both \textit{pusher} and \textit{puller} swimmers : one can notice that the larger the magnetic field, the stronger the \textit{motor-brake} effect. For a given $Pe_{m}$, the effect is maximal for $\alpha=45^{\circ}$ and $\alpha=135^{\circ}$ for which the extensional and compression axis of the flow created by the particles are aligned (\textit{motor}) or perpendicular (\textit{brake}) to the ones of the imposed simple shear flow. Similarly, for $\alpha=0,\ 90\mathrm{\ and}\ 180^{\circ}$, the actuated stress vanishes because the magnetic swimmers shear the fluid in the orthogonal direction to the imposed shear. \\
\indent To investigate more quantitatively the \textit{motor-brake} effect, we compute an analytical asymptotic expression for the active and drag dimensionless stresses to leading orders in $Pe_{m}$ and $Pe_{H}$. As mentioned above, the magnetic and diffusive stresses do compensate each other in the limit of low $Pe_{H}$ and $Pe_{m}$, such that their combined contribution vanishes in the asymptotic expansion of the stress. Note that, in other limits which are not investigated by this paper, the contributions of these stresses do not compensate and must be considered explicitly in the total stress balance. We restrict the expansion of the distribution function $\Psi(\theta,\phi)$ to the first two-spherical harmonics : 
\begin{equation}
\Psi(\theta,\phi)=\dfrac{1}{4\pi}\left(\displaystyle{\sum_{n=0}^{2}\sum_{l=0}^{n}A_{n}^{l}P_{n}^{l}(\cos\theta)\cos l\phi}+\displaystyle{\sum_{n=1}^{2}\sum_{l=1}^{n}B_{n}^{l}P_{n}^{l}(\cos\theta)\sin l\phi}\right)
\end{equation}
where $A_{n}^{l}$ and $B_{n}^{l}$ are the coefficients of the expansion, solutions of the Fokker-Planck equation~(\ref{developed_FP_eq}).
In this expansion, the spherical harmonics which contribute dominantly to the stress terms $<p_{x}p_{y}>$ and $<p_{x}^2p_{y}^2>$ are respectively $\Psi_{2}^{2}(\theta,\phi)=B_{2}^{2}P_{2}^{2}(\cos\theta)\sin2\phi/(4\pi)$ and $\Psi_{0}^{0}(\theta,\phi)=A_{0}^{0}/(4\pi)=1/(4\pi)$. Injecting $\Psi(\theta,\phi)$ into the Fokker-Planck equation~(\ref{developed_FP_eq}), we show that $B_{2}^{2}=1/12\left( \sin(2\alpha)Pe_{m}^2 +Pe_{H}\right)$ and we obtain an asymptotic scaling expression for the particle shear stress:
\begin{equation}
\widetilde{\Sigma}_{p}=\dfrac{1}{30}\left[\epsilon\sin(2\alpha)Pe_{m}^{2}+\left(\dfrac{1}{\mathcal{A}}+\epsilon\right)Pe_{H}\right] + o(Pe_{H},Pe_{m}^{2})
\label{asymptotic_particle_stress}
\end{equation}
which can be written in the form $\widetilde{\Sigma}_{p}=\widetilde{\Sigma}_{0}+\eta_{p}Pe_{H}$, where $\widetilde{\Sigma}_{0}=\dfrac{\epsilon}{30}\sin(2\alpha)Pe_{m}^{2}$ is the actuated stress and $\eta_{p}=\dfrac{1}{30}\left( \dfrac{1}{\mathcal{A}}+\epsilon \right)$ is the particle-borne viscosity contribution. A cross-over between the linear and the actuated stress regimes is observed at 
$Pe_{H}^{\mathrm{*}}=\left\lvert\dfrac{\sin(2\alpha)(Pe_{m})^{2}}{1/\mathcal{A}+\epsilon}\right\rvert $.
A master curve for this asymptotic scaling limit can be obtained rescaling the particle stress by $\left\lvert\widetilde{\Sigma}_{0}\right\rvert$  and the hydrodynamic Peclet number by the cross-over value $Pe_{H}^{*}$ i.e. : $\widetilde{Pe}_{H}=Pe_{H}/Pe_{H}^{\mathrm{*}}=Pe_{H}/\left\lvert\dfrac{\sin(2\alpha)(Pe_{m})^{2}}{1/\mathcal{A}+\epsilon}\right\rvert$. The expression for this master curve is then :
\begin{equation}
\dfrac{\widetilde{\Sigma}_{p}}{\lvert\widetilde{\Sigma}_{0}\rvert}=\epsilon \ \mathrm{sign}(\sin(2\alpha)) + \widetilde{Pe}_{H}\ \mathrm{sign}(1/\mathcal{A}+\epsilon)
\label{equation_rescaling}
\end{equation}
To test the validity of the scaling expression, we propose to compare the results stemming from the full numerical solution of the Fokker-Planck equation yielding  $\Psi(\theta,\phi)$ and the particle stress $\widetilde{\Sigma}_{p}$ (equation~(\ref{nondim_particle_stress})), with the analytical expression for the master curve (equation~(\ref{equation_rescaling})). From the computation of $\widetilde{\Sigma}_{p} (Pe_{H})$ at different $Pe_{m}$, we extract the value of $\widetilde{\Sigma}_{0}$ and $Pe_{H}^{*}$ using an affine fit in $Pe_{H}$ (see figure~\ref{illustration_fit}(a)). We then construct the master curve of figure~\ref{collapse_plot} where each data is a symbol : we indeed observe a collapse of the full numerical solution of equation~(\ref{nondim_particle_stress}) onto the master equation~(\ref{equation_rescaling}) for a wide range of parameters. This scaling law is no longer valid at large $Pe_{m}$, beyond the validity limit of the expansion done in equation~(\ref{asymptotic_particle_stress}), which we determine by computing the deviation of $\widetilde{\Sigma}_{p} (Pe_{H})$ from an affine fit (see figure~\ref{illustration_fit}(a)). The validity domain is reported figure~\ref{diagramme_validite_Bingham}(b).\\
\indent The total shear stress of the suspension (including the contributions of both the suspending fluid and the magnetic micro-swimmers) is then, in its dimensional version~:
\begin{equation}
\Sigma_{xy}=\eta_{s}\dot{\gamma}+n\sigma_{0}\widetilde{\Sigma}_{p}
\label{total_dimensioned_part_stress}
\end{equation}
where $n=\Phi/\mathcal{V_{B}}=6\Phi/(\pi a^{2}L)$ is the volume density of active particles in the fluid, $\Phi$ is the particle volume fraction and $\mathcal{V_{B}}$ the volume of a single ellipsoidal particle. Injecting the asymptotic scaling~(\ref{asymptotic_particle_stress}) of $\widetilde{\Sigma}_{p}$ into equation~(\ref{total_dimensioned_part_stress}), we obtain the dimensional total shear stress exerted on the suspension~:
\begin{equation}
\begin{array}{ll}
\Sigma_{xy}= &\underbrace{\epsilon\dfrac{n\sigma_{0}}{30} \sin(2\alpha)Pe_{m}^{2}}_{\Sigma_{0}} \\
& \\
&+\underbrace{\left( \eta_{s} +\dfrac{n\sigma_{0}}{30D_{r}}\left(\dfrac{1}{\mathcal{A}}+\epsilon\right) \right)}_{\mathrm{\eta_{\mathrm{eff}}}}\dot{\gamma}
\end{array}
\label{Bingham_form}
\end{equation}
This constitutive relation generalizes the result of Saintillan \cite{Saintillan2010}. It contains the dimensional \textit{actuated stress} $\Sigma_{0}\equiv n\sigma_{0}\widetilde{\Sigma}_{0}$ described above and a linear dependance with the shear rate $\dot{\gamma}$, defining an effective viscosity of the suspension~$\eta_{\mathrm{eff}}\equiv \partial \Sigma_{xy}/\partial \dot{\gamma}$. Remarkably, the magnetic field angle can be chosen so that the actuated shear stress becomes negative for both pusher and puller swimmers. When the actuated stress dominates, the swimming power of the bacteria transferred to the fluid induces a shear of the suspension which can be oriented in the same direction as the imposed shear (\textit{motor state}) or in the opposite direction (\textit{brake state}). \\
\indent An other feature of equation~(\ref{Bingham_form}) is that $\eta_{\mathrm{eff}}$ is identical to the effective viscosity of non-magnetic micro-swimmers suspensions, derived in~\cite{Saintillan2010} for instance. This comes from the fact that, in the range of validity of equation~(\ref{Bingham_form}), there is no coupling between magnetic and hydrodynamic terms in the stress. However, in other ranges of parameters, non-trivial couplings between these terms change the effective viscosity of the suspension. \\
\indent Note that, for $Pe_{m}\gg 1$, i.e. out from the validity limit of expression~(\ref{Bingham_form}), the actuated stress exhibits a saturation. Indeed, in this limit, the micro-swimmers are mainly aligned in the magnetic field direction and deliver collectively the maximum shear allowed by their swimming energy. Then, the maximal intensity of the \textit{motor-brake} effect is reachable by achieving the limit $Pe_{m}\gg 1$, which is confirmed by the numerical computation of the actuated stress (see inset of figure~\ref{collapse_plot}).

\section{IV. GENERALITY OF THE \textit{MOTOR-BRAKE} EFFECT}

In this section, we show that the properties of the \textit{motor-brake} effect are not specific to the hypothesis of our model. \\
\indent First, the rheological response of SS swimmers is the same as the one of NS swimmers. By symmetry, SS corresponds to NS after a re-orientation of the magnetic field by an angle $\pi$. Thus, if the NS swimmers are in \textit{motor} state, the SS will also be in \textit{motor} state and \textit{vice-versa}.

\begin{figure}[h!]
\centering
\includegraphics[scale=.60]{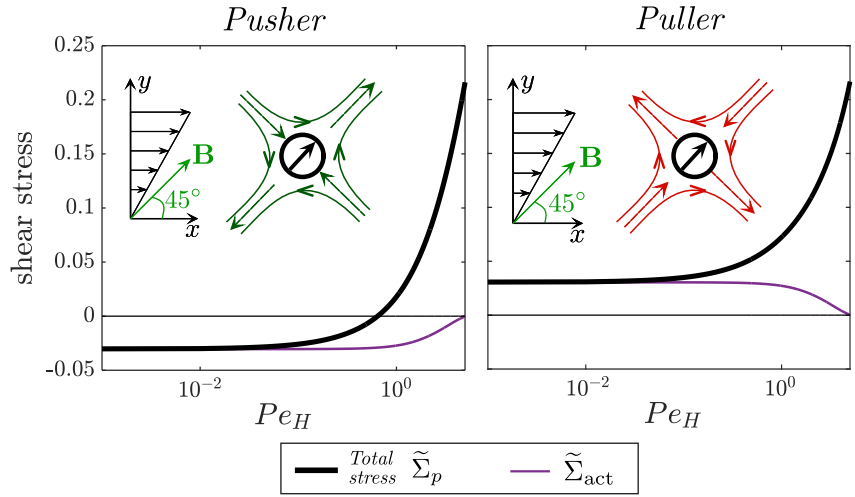}
\caption{Computation of the particle shear stress and active stress for spherical micro-swimmers. The \textit{motor-brake} effect is recovered also for these kind of particles and the value of the \textit{actuated stress} is identical to the one of rod-shaped particles.}
\label{puller_pusher_spherical}
\end{figure}

We also investigate the role played by the particle geometry. More specifically, we analyzed the case of a purely spherical particle ($\beta=0$). In this case, the kinematic equation is changed :
$\dot{\mathbf{p}}=(\overline{\overline{\mathbf{I}}}-\mathbf{p}\mathbf{p})\overline{\overline{\mathbf{\Omega}}}\cdot\mathbf{p}+\mathbf{\Omega_{m}}\times\mathbf{p}=\left(\mathbf{\Omega}+\mathbf{\Omega_{m}}\right)\times\mathbf{p}$,
where $\mathbf{\Omega}$ is the fluid vorticity vector. The corresponding change in the Fokker-Planck equation~(\ref{developed_FP_eq}) concerns the operator $\Gamma_{\textrm{shear}}(\Psi)=\dfrac{1}{2}\dfrac{\partial \Psi}{\partial \phi}$. Concerning the stresses, the expressions are also different from the elongated rod case. The dimensional particle stress then reads :
\begin{equation}
\overline{\overline{\mathbf{\Sigma}}}_{p}=\dfrac{5}{6}n\xi_{r}\overline{\overline{\mathbf{E}}}+n\sigma_{0}\left[<\mathbf{p}\mathbf{p}>-\dfrac{\overline{\overline{\mathbf{I}}}}{3}\right]+\dfrac{nmB}{2}\left[ <\mathbf{p}>\mathbf{b} -\mathbf{b}<\mathbf{p}> \right]
\end{equation}
In this stress, the contribution of diffusive processes are zero due to the symmetry of the particle. Only remain the stress corresponding to the friction on the particle body, the active stress and the traceless magnetic stress (it is antisymmetric here due to the spherical geometry of the particle, see~\cite{Kim1994}).\\
\indent On figure~\ref{puller_pusher_spherical} are some numerical results obtained for both pusher and puller spherical swimmers. The phenomenology is the same as for elongated particles, meaning the appearence of a constant shear stress at low shear rate (or $Pe_{H}$). Moreover, the value of this constant, identical to the one of rod-shaped particles, is solely due to the active stress and the value and the sign of this constant depends on $Pe_{m}$, $\alpha$ and $\epsilon$, in the same way as for elongated swimmers. This shows that the \textit{motor-brake} effect is general for various kinds of particle shapes. The main quantitative difference between the case of spherical and elongated particles is the effective viscosity of the suspension which is always positive for spherical particles, as already described before for non-magnetic swimmers~\cite{Saintillan2010}.

\section{V. ORDERS OF MAGNITUDE OF THE EFFECT AND CONCLUSION}

In order to test whether the \textit{motor-brake} effect can be significant experimentally, let us take the example of a suspension of magnetotactic bacteria of dimensions $L=5$ $\mu$m, $a=1$ $\mu$m  at a volume fraction of 1\% (i.e. of number density $n\simeq 10^{16}$ m$^{-3}$). The force dipole magnitude of an active swimmer moving with a velocity of $10$ $\mu$m.s$^{-1}$ is typically $\sigma_{0}=10^{-18}$ J ~\cite{Drescher2011}. From experiments on magnetotactic bacteria using the \textit{Magnetospirillum gryphiswaldense} MSR-1 strain, we get values of $m\simeq 10^{-16}$ A m$^{-2}$ (see also \cite{Reufer2014,Nadkarni2013}); $D_{r}=1$ s$^{-1}$, $t_{H}=0.1$ s (typical time for the bacteria to move over its size). The activity number $\mathcal{A}$ is then typically equal to 10, and for a magnetic field $B=1$ mT, $Pe_{m}=1$. The magnetic field orientation is chosen to be at $\alpha=135^{\circ}$. It is then possible to evaluate the $Pe_{H}$ below which $\Sigma_{xy}<0$, using equation~(\ref{Bingham_form}). We obtain $Pe_{H}\simeq 1$ which corresponds to $\dot{\gamma}\simeq 1$ s$^{-1}$. The value obtained for $Pe_{H}$ remains in the domain of validity of equation~(\ref{Bingham_form}). The corresponding shear rate magnitude can be reached by known rheometry~\cite{Gachelin2014,Lopez2015}, meaning that this effect could indeed be observed experimentally. Note that published experimental setups would allow to test the effect in the conditions akin to the model~\cite{Cheng2011}. Moreover, one can estimate a numerical value for the maximum shear stress available from the system at low shear rate, which is $n \sigma_{0}\sim 10^{-2}-10^{-1}$ Pa. This value needs to be compared to typical pressure loss in microfluidics~: $10^{-1}$ Pa is needed to flow water in a cylindrical channel of 1cm-length and 100$\mu$m-radius at 1nL.s$^{-1}$, which is of the same order of magnitude. This indicates that the effect could be used to control microfluidic flows.

For standard rheo-magnetic suspensions, a negative-viscosity effect was found in response to oscillatory magnetic fields~\cite{Bacri1995} or adding a constant torque on non-colloidal particles~\cite{Jibuti2012}. The \textit{motor-brake} effect derived here is very different conceptually. First, it relies on the activity of magnetic swimmers under a constant magnetic field. Second, the actuation of a  constant negative shear stress at low shear rate (\textit{motor state}) is new in rheology. The tunability of the \textit{motor} and \textit{brake} states for such suspensions could open the way to several practical applications, as direct flow control in microfluidic devices or energy harvesting to build microscopic motorized systems.

\begin{acknowledgments}
 We acknowledge the support of the ANR-2015 ``BacFlow'', a critical reading of the manuscript by Dr. Laurette Tuckerman and scientific discussions with Prof. Anke Lindner. 
\end{acknowledgments}
\vspace{-4 mm}


\begin{thebibliography}{}
\bibitem{Lauga2009} E.Lauga \& T.R. Powers, \textit{The hydrodynamics of swimming microorganisms}, Rep. Prog. Phys., \textbf{72}, 096601 (2009).

\bibitem{Paxton2004} W. F. Paxton et al., \textit{Catalytic nanomotors: autonomous movement of striped nanorods}, J. Am. Chem. Soc., 126, \textbf{13}, 424 (2004).

\bibitem{Maass2016} C. C. Maass, C.Krüger, S. Herminghaus \& C. Bahr, \textit{Swimming droplets}, Annual Review of Condensed Matter Physics, \textbf{7}, 171-193  (2016).

\bibitem{Dreyfus2005} R. Dreyfus et al., \textit{Microscopic artificial swimmers}, Nature, \textbf{437}, 862 (2005).
\bibitem{Tierno2008} P.Tierno, \textit{Magnetically actuated colloidal microswimmers}, J. Phys. Chem. \textbf{B}, 112, 16525 (2008).
\bibitem{Martel2012} S.Martel, \textit{Bacterial microsystems and microrobots}, Biomed Microdevices, \textbf{14}, 1033 (2012).

\bibitem{Williams2014} Williams B. J., Anand S. V., Rajagopalan J. \& Saif, M. T. A., \textit{A self-propelled biohybrid swimmer at low Reynolds number}, Nature Com., \textbf{5}, 1–8 (2014).

\bibitem{Koch2011}D.L Koch \& G. Subramanian, \textit{Collective hydrodynamics of swimming microorganisms: living fluids}, Annual Review of Fluid Mechanics, {\bf 43}, 637--659 (2011).
\bibitem{Marchetti2013} M.C Marchetti {\it et al.}, \textit{Hydrodynamics of soft active matter}, Rev. Mod. Phys., {\bf 85}, 1143 (2013).

\bibitem{Hatwalne2004} Y. Hatwalne, S. Ramaswamy, M. Rao \& R. Aditi Simha, \textit{Rheology of active-particle suspensions}, Phys. Rev. Lett., {\bf 92,} 118101 (2004).

\bibitem{Toner2005}J. Toner, Y. Tu \& S. Ramaswamy, \textit{Hydrodynamics and phases of flocks}, Ann. Phys., \textbf{318}, 170 (2005).

\bibitem{Wu2000} X.-L. Wu et al., \textit{Particle diffusion in a quasi-two-dimensional bacterial bath}, Phys. Rev. Lett., \textbf{84}, 3017 (2000).

\bibitem{Sokolov2009}S. Sokolov \& I.S. Aranson, \textit{Reduction of viscosity in suspension of swimming bacteria}, Phys. Rev. Lett., {\bf 103,} 148101 (2009).

\bibitem{Rafai2010} S. Rafa\"\i, L. Jibuti \& P. Peyla, \textit{Effective viscosity of microswimmer suspensions}, Phys. Rev. Lett., \textbf{104}, 098102 (2010).

\bibitem{Leptos2009}K. C. Leptos et al., \textit{Dynamics of enhanced tracer diffusion in suspensions of swimming eukaryotic microorganisms}, Phys. Rev. Lett., \textbf{103}, 198103 (2009).

\bibitem{Sokolov2010} A. Sokolov et al., \textit{Swimming bacteria power microscopic gears}, Proc. Natl. Acad. Sci. U.S.A., {\bf 107}, 969 (2010).

\bibitem{DiLeonardo2010} R. Di Leonardo {\it et al.}, \textit{Bacterial ratchet motors}, Proc. Natl. Acad. Sci. U.S.A., {\bf 107}, 9541 (2010).
\bibitem{Mino2011} Mino G \textit{et al}, \textit{Enhanced diffusion due to active swimmers at a solid surface}, Phys. Rev. Lett, \textbf{106} 048102 (2011).

\bibitem{Rusconi2014}R. Rusconi, J.S. Guasto \& R. Stocker, \textit{Bacterial transport suppressed by fluid shear}, Nature Physics, {\bf 10}, 212-217 (2014).
\bibitem{Gachelin2014}J. Gachelin, A. Rousselet, A. Lindner \& E. Clement, \textit{Collective motion in an active suspension of Escherichia coli bacteria}, New Journal of Physics, {\bf 16}, 025003 (2014).

\bibitem{Lopez2015} Lopez H.M., J. Gachelin, C. Douarche, H.Auradou, E. Clement, \textit{Turning bacteria suspensions into superfluids}, Phys. Rev. Lett., \textbf{115}, 028301 (2015).

\bibitem{Brady2017}S.C. Takatori \& J.F. Brady, \textit{Superfluid Behavior of Active Suspensions from Diffusive Stretching}, Phys. Rev. Lett., {\bf 118}, 018003 (2017).

\bibitem{Cates2008}M.E. Cates, S.M Fielding, D. Marenduzzo, E. Orlandini \& J.M. Yeomans, \textit{Shearing active gels close to the isotropic-nematic transition}, Phys.Rev. Lett., {\bf 101}, 068102 (2008).

\bibitem{Giomi2010}L. Giomi, T.B. Liverpool \& M.C. Marchetti, \textit{Sheared active fluids: Thickening, thinning, and vanishing viscosity}, Phys. Rev. E, {\bf 81}, 051908 (2010).

\bibitem{Uebe2016} R.Uebe, D.Schüler, \textit{Magnetosome biogenesis in magnetotactic bacteria}, Nature Reviews Microbiology, \textbf{14}, 621–637 (2016).

\bibitem{Reufer2014} M. Reufer \textit{et al.}, \textit{Switching of swimming modes in Magnetospirillium gryphiswaldense}, The Biophysical Journal, {\bf 106}, 37-46 (2014)

\bibitem{Waisbord2016} N. Waisbord, C. T. Lef\`evre, L. Bocquet, C. Ybert \& C. Cottin-Bizonne, \textit{Destabilization of a flow focused suspension of magnetotactic bacteria}, Phys. Rev. Fluids, {\bf 1} (2016)

\bibitem{Jeffery1922} G.B. Jeffery, \textit{The motion of ellipsoidal particles immersed in a viscous fluid}, Proc. R. Soc. London, Ser. A, \textbf{102}, 161 (1922).

\bibitem{Bretherton1962} F.P. Bretherton, \textit{The motion of rigid particles in a shear flow at low Reynolds number}, J.Fluid Mech., \textbf{14}, 284 (1962).

\bibitem{Satoh2001} A. Satoh, \textit{Rheological Properties and Orientational Distributions of Dilute
Ferromagnetic Spherocylinder Particle Dispersions}, Journal of Colloid and Interface Science, \textbf{234}, 42533 (2001).

\bibitem{Haines2009} Haines B.M. {\it et al.}, \textit{Three-dimensional model for the effective viscosity of bacterial suspensions}, Phys. Rev. E ,{\bf 80}, 041922 (2009).

\bibitem{Saintillan2010}D. Saintillan, \textit{The dilute rheology of swimming suspensions: A simple kinetic model}, Exp. Mech., {\bf 50}, 125 (2010).

\bibitem{Brenner1974}H. Brenner, \textit{Rheology of a dilute suspension of axisymmetric Brownian particles}, Int. J. Multiphase Flow, {\bf 1}, 195-341 (1974).

\bibitem{Jansons1983}Jansons KM, \textit{Determination of the constitutive equations for a magnetic fluid}, J. Fluid. Mech., \textbf{137}, 187-216 (1983).

\bibitem{Rosensweig} R. E. Rosensweig, \textit{Ferrohydrodynamics}, Cambridge University Press {\bf }(1985).

\bibitem{Kim1994} S.R. Strand \& S. Kim, \textit{Dynamics and rheology of a dilute suspension of dipolar nonspherical particles in an external field: Part 1. Steady shear flows}, Rheologica Acta, {\bf 31}, 94-117 (1992).

\bibitem{Leal&Hinch1976}E. J. Hinch \& L.G. Leal, \textit{Constitutive equations in suspension mechanics. Part 2. Approximate forms for a suspension of rigid particles affected by Brownian rotations}, Journal of Fluid Mechanics, {\bf 76}, 187-208 (1976).

\bibitem{Nadkarni2013} R. Nadkarni, S. Barkley \& C. Fradin, \textit{A comparison of methods to measure the magnetic moment of magnetotactic bacteria through analysis of their trajectories in external magnetic fields}, PLOS ONE, {\bf 8}, 12 (2013).

\bibitem{Cheng2011} X. Cheng, J. H. McCoy, J. N. Israelachvili \& I. Cohen, \textit{Imaging the microscopic structure of shear thinning and thickening colloidal suspensions}, Science, {\bf 333}, 1276-1279 (2011).

\bibitem{Bacri1995} J.-C. Bacri, R. Perzynski, M. I. Schliomis \& G. I. Burde, \textit{“Negative-viscosity” effect in a magnetic fluid}, Phys. Rev. Lett., {\bf 75} 2128 (1995).

\bibitem{Jibuti2012}L. Jibuti, S. Rafai \& P. Peyla, \textit{Suspensions with a tunable effective viscosity: a numerical study}, Journal of Fluid Mechanics, \textbf{693}, 345-366 (2012). 

\bibitem{Drescher2011}K. Drescher, J. Dunkel, L. H. Cisneros, S. Ganguly \& R. E. Goldstein,  \textit{Fluid dynamics and noise in bacterial cell–cell and cell–surface scattering}, PNAS, \textbf{108}, 27, 10940–10945 (2011).

\bibitem{Saintillan2018}D. Saintillan, \textit{Rheology of Active Fluids}, Annual Review of Fluid Mechanics, \textbf{50}, 563-592 (2018).

\end{thebibliography}
\end{document}